\documentclass[10pt]{article}

\usepackage{amsmath}
\usepackage{amssymb}

\usepackage{graphicx}
\usepackage{epsfig}

\usepackage{cite}

\usepackage{color}
\usepackage[labelfont=bf,labelsep=period,justification=raggedright]{caption}

\makeatletter
\renewcommand{\@biblabel}[1]{\quad#1.}
\makeatother

\date{}

\begin{document}

% Title must be 150 characters or less
\begin{flushleft}
{\Large
\textbf{Extraordinary Sex Ratios: Cultural Effects on Ecological Consequences}
}
% Insert Author names, affiliations and corresponding author email.
\\
Ferenc Moln\'{a}r Jr.$^{1}$, Thomas Caraco$^{2}$, Gyorgy Korniss$^{1,\ast}$
\\
\bf{1} Department of Physics, Applied Physics, and Astronomy, Rensselaer Polytechnic Institute, Troy, NY, USA
\\
\bf{2} Department of Biological Sciences, University at Albany, Albany, NY, USA
\\
$\ast$ E-mail: korniss@rpi.edu
\end{flushleft}

% Please keep the abstract between 250 and 300 words
\section*{Abstract}
We model sex-structured population dynamics to analyze pairwise
competition between groups differing both genetically and
culturally.  A sex-ratio allele is expressed in the heterogametic
sex only, so that assumptions of Fisher's analysis do not apply.
Sex-ratio evolution drives cultural evolution of a group-associated
trait governing mortality in the homogametic sex. The two-sex
dynamics under resource limitation induces a strong Allee effect
that depends on both sex ratio and cultural trait values.  We
describe the resulting threshold, separating extinction from
positive growth, as a function of female and male densities.  When
initial conditions avoid extinction due to the Allee effect,
different sex ratios cannot coexist; in our model, greater female
allocation always invades and excludes a lesser allocation.  But the
culturally transmitted trait interacts with the sex ratio to
determine the ecological consequences of successful invasion.
The invading female allocation may permit population persistence at
self-regulated equilibrium.  For this case, the resident culture may
be excluded, or may coexist with the invader culture.  That is, a
single sex-ratio allele in females and a cultural dimorphism in male
mortality can persist; a low-mortality resident trait is maintained
by father-to-son cultural transmission.  Otherwise, the successfully
invading female allocation excludes the resident allele and culture,
and then drives the population to extinction via a shortage of
males. Finally, we show that the results obtained under homogeneous
mixing hold, with caveats, in a spatially explicit model with local
mating and diffusive dispersal in both sexes.

%\pacs{87.23.Cc,  %Population dynamics and ecological pattern formation
%      87.23.Kg,  %Dynamics of evolution
%      82.40.Ck   %Pattern formation in reactions with diffusion
%}

\section*{Introduction}
Since Fisher's \cite{Fisher_1930} classic insight, sex-ratio
evolution \cite{Shaw_1953,Eshel_1975,Karlin_1986} and the impact of
a given sex ratio on ecological dynamics
\cite{Caswell_1986,Ashih_2001,Engen_2003,Miller_2011} have remained
central issues in population biology.  Fisher \cite{Fisher_1930}
noted that neither sex should be rarer at evolutionary equilibrium,
a consequence of frequency-dependent selection.  That is, equal
investment of reproductive effort in the two sexes --- commonly
implying a sex ratio close to unity --- can be evolutionarily stable
\cite{Charnov_1982}.

Hamilton \cite{Hamilton_1967} studied sex ratios departing
significantly from unity, emphasizing that Fisher's argument does
not apply when a sex-linked gene controls sex ratio at birth.  In
particular, if a gene governing sex ratio occurs in the
heterogametic sex only (females in the $ZW$ system, and males in the
$XY$ system), the gene's fitness depends only on the number of
heterogametic offspring produced.  The frequency of such a gene may
advance rapidly, endangering population persistence
\cite{Jaenike_2001,Tainaka_EPL2006}.  That is, a biased sex ratio
can leave members of the more common sex without mates; the
consequent ``marriage squeeze'' \cite{Caswell_1986} may lead to
population decline \cite{Hamilton_1967,West_2009}.  Equivalently, an
Allee effect (dependent on the density of each sex) can limit the
degree of sex-ratio bias, for given total density, capable of
averting direct decline to extinction
\cite{Ashih_2001,BBB_AmNat2001,BB_JTB2002,BerecTREE_2006}. Our study
supposes that an extraordinary sex ratio's ecological consequence,
population persistence or extinction, depends on interaction with a
culturally inherited trait.

Cultural traits may enforce a between-sex mortality difference
\cite{Kumm_1994}.  In certain human cultures, infanticide and
neglect increase female mortality
\cite{Dickemann_1975,Hausfater_1984}; Laland \emph{et al}.
\cite{Laland_1995} assume that these cultural traits are transmitted
vertically, \emph{i.e}., parent to offspring. In other species,
vertical cultural transmission clearly causes between-sex
differences in habitat choice, tool use or foraging behavior, but
their relationships to sex-specific mortality rates are unknown
\cite{Diamond_1987,Lonsdorf_2004,Krutzen_2005,Slagsvold_2011}. Our
models explore how a cultural trait influencing male mortality might
govern the ecological consequences of sex-ratio evolution.  We treat
sex ratio as a sex-linked genetic trait, and restrict cultural
transmission to the vertical case \cite{LCS_1981}.  Our two-sex
population dynamics assumes competition for a growth-limiting
resource; competition generates a strong Allee effect.  Within a
group, each female carries the same sex-ratio allele, and each male
experiences the same mortality rate; parameters differ between
groups.  Resource competition is preemptive; each group has the same
niche \cite{Amarasekare_2003,Allstadt_2009,Going_2009}.

Our approach assumes pairwise competition between resident and
invader groups, where group refers to population structure, not the
level of selection. In Sober's \cite{Sober_1984} terminology, we
associate properties driving selection with groups, and associate
the objects of selection with individuals --- individual females in
this case.
The resident group (sex ratio, male mortality culture)
rests at ecological equilibrium, and we ask if a rare, different
group can invade the resident.  Our results for  invasion,
extinction and (cultural) coexistence indicate how resource
competition, cultural variation and sex-ratio evolution interact.
Ecological invasion often has a distinctly spatial character
\cite{KC_JTB2005,OBYKAC_TPB2006}.  Therefore, we extend our model
beyond the assumption of homogeneous mixing, and introduce spatial
detail by analyzing the model's reaction-diffusion extension.

\section*{Methods}
\subsection*{General assumptions}
In birds (and butterflies) sex determination follows the $ZW$
system. $W$ is the sex-determining chromosome;  females are $ZW$,
and males are $ZZ$ \cite{Hastings_1994}.  Our model assumes that the
$W$ chromosome carries an allele fixing the sex ratio among that
female's offspring.  The sex linkage means that a female inherits
her mother's sex ratio, and the sex-ratio gene never occurs in
males.  Hence, the fitness of the sex-ratio allele (of any gene on
the $W$ chromosome) is advanced only through production of daughters
\cite{Hamilton_1967}. To focus our discussion accordingly, we model
the ``female ratio,'' the proportion of a female's offspring born
female. Females of a single group carry the same sex-ratio allele.

The assumption of sex-linkage might seem restrictive.  However, in a
number of bird species, individual females shed Z-chromosome and
W-chromosome bearing eggs non-randomly
\cite{Appleby_1997,Komdeur_1997}.  The observed variation in sex
ratio among females may reflect facultative plasticity
\cite{Dijkstra_1990}, but could generate some of the
population-dynamic consequences of sex ratio that we model.

All members of a given group
share a vertically transmitted cultural norm that governs male
behavior which, in turn, fixes the male mortality rate for that
group.  Females of different groups share the same mortality rate.
Hence, for simplicity, we assume a female adopts her mother's
culture.  If both parents belong to the same group, their son
faithfully acquires the parental culture.  When parents of different
cultures (groups) mate, a son acquires one or the other culture,
each with probability $1/2$.

To address competition between groups, we envision a resident group
(a single female ratio and a single male mortality rate) at
ecological equilibrium in a resource-limited environment.  We then
introduce (\emph{via} demic/genetic migration) a small inoculum of
an invader group.  The resident and the rare invader differ in
female ratio and ordinarily differ in male mortality.  The
competitive dynamics proceeds to ecological equilibrium.  If the
rare female-ratio allele has positive growth, it will drive change
in culture.  Since individuals mate randomly, extinction of a
group's female-ratio allele need not always imply loss of the
associated cultural trait.  However, loss of a cultural mortality
trait implies that the associated female-ratio allele has been
excluded competitively.

Our population dynamics differs from models for gene-culture
coevolution where different alleles and cultural traits directly
affect each other's evolution \cite{Laland_1995}.  Our model's
cultural trait directly influences the resident's population density
and the invader's growth rate when rare; female ratios and male
mortalities interactively drive the invader's dynamics.  We do not
assume functional dependence between the genetic and cultural
traits.  Rather, we evaluate consequences of the feasible range of
male-mortality rate combinations for the entire range of female
ratio combinations (resident and invader).

\subsection*{Mathematical model}
Consider two-sex population growth with two female ratio/male
mortality groups; the groups allow us to model resident-invader
differences.  When a female of group $i$ $(i = 1,\: 2)$ reproduces,
the resulting offspring is female with probability $\theta_{i}$, and
male with probability $(1-\theta_{i})$, independently of the group
of the male with whom she mates.  $\theta_{i}$ is the female ratio
for group $i$, transmitted faithfully from mother to daughter.
Different groups, by definition, differ in female ratio.  All
females have the same mortality rate, $\mu_{\rm f}$.

A male's group specifies his mortality rate, $\mu_i$ $(i = 1,\: 2)$.
If male mortality exceeds the rate for females, $\mu_1 , \mu_2 >
\mu_{\rm f}$.  But we do not exclude the case where the female
mortality exceeds one or both male rates.  If both parents belong to
the same group, each male offspring has that group's mortality rate,
acquired by vertical cultural transmission.  If a male's parents
belong to different groups, the male acquires mortality rate $\mu_i$
with probability $\frac{1}{2}$.

$F_{i}$ and $M_{i}$ represent the global density of females and
males, respectively, of group $i$.  All individuals require the same
resources, so that population growth at larger densities will
self-regulate.  The preceding assumptions imply the following
dynamics under homogeneous mixing (or ``mean-field"):
\begin{eqnarray}
\partial_{t} F_{1} & = & \theta_{1} \left(1-N\right) F_{1} \left(M_{1}+M_{2}\right) - \mu_{\rm f} F_{1} \nonumber \\
\partial_{t} M_{1} & = & \left(1-N\right) \left[\left(1-\theta_{1}\right) F_{1} \left(M_{1}+\frac{M_{2}}{2}\right)+\left(1-\theta_{2}\right) F_{2}  \left(\frac{M_{1}}{2}\right)\right] - \mu_{1}M_{1} \nonumber \\
\partial_{t} F_{2} & = & \theta_{2} \left(1-N\right) F_{2} \left(M_{1}+M_{2}\right) - \mu_{\rm f} F_{2} \nonumber \\
\partial_{t} M_{2} & = & \left(1-N\right) \left[ \left( 1-\theta_{2} \right) F_{2} \left(\frac{M_{1}}{2}+M_{2}\right) + \left(1-\theta_{1}\right) F_{1}  \left(\frac{M_{2}}{2}\right) \right] - \mu_{2} M_{2} \;,
\label{eq:model4}
\end{eqnarray}
where $N = F_{1}+M_{1}+F_{2}+M_{2}$ is total global density; $0 \leq
N \leq 1$.  Males encounter females as a mass-action process,
modeling random mating \cite{LCS_1981,BBB_AmNat2001}; more
complicated assumptions about pair formation suggest different
``marriage functions'' \cite{Miller_2011}.  The fraction of matings
that reproduce successfully equals the unoccupied fraction of the
environment, $(1-N)$.  Below we take group 1 as the resident, and
identify group 2 as the (initially rare) invader.

If only a single group occupies the environment, the equations
reduce to those studied by Tainaka et al. \cite{Tainaka_EPL2006}:
\begin{eqnarray}
\partial_{t} F & = & \theta \left(1-M-F\right) F M - \mu_{\rm f} F \nonumber \\
\partial_{t} M & = & \left(1-\theta\right) \left(1-M-F\right) F M - \mu_{\rm m} M \;.
\label{eq:model2}
\end{eqnarray}
The authors focused on the symmetric case, $\mu_{\rm f}=\mu_{\rm
m}$. An important feature of this model is that the cubic dynamics
produces a strong Allee effect \cite{BB_JTB2002,BerecTREE_2006}.
That is, there exists a threshold for the initial population
density, below which growth is necessarily negative, and extinction
must follow \cite{BBB_AmNat2001}.  The single-group model serves as
the starting point of our analysis.  In particular, initial
conditions of our competition dynamics will depend on the stable,
non-trivial fixed point of the single-group model (corresponding to
positive equilibrium densities for females and males of group  1).

\subsection*{Analytic and numerical methods} %%%%%%%%%%%%%%%%%%%%%%%%%%%%%%%%%%%%%%%%%%%%%%%%%%%%%%%

We assume that the population dynamics is fast compared to the time
scale of immigration (invasion of new gene-culture groups).  Then
female ratio should evolve through a series of successful invasions
of populations resting at demographic equilibrium.  Therefore, we
obtained the fixed points (stationary solutions) of
Eqs.~(\ref{eq:model4}) analytically (see Supporting Information S1), and we used
numerical integration to analyze their local stability.

Since this study employs extensive numerical integration, we justify
our choice of an ordinary differential equation (ODE) solver.
Equations~(\ref{eq:model4}) are strongly coupled and may become
stiff, a challenge to the solver.  Speed is another important factor
because we mapped the entire parameter space of the model, which
requires a very large amount of computation. We chose the explicit
fourth-order Runge-Kutta method \cite{numrecip-rk4}, which gives the
precision we require. We utilized adaptive time stepping to avoid
problems with any potential stiffness, and to increase integration
speed when the slopes of the densities were small.  Since we are
interested in stationary solutions of the equations, the stopping
condition for the integration specifies that all numerical
derivatives are smaller than a predetermined limit:
\begin{equation}
\frac{\Delta A}{\Delta t} < \epsilon,\; A \in \left\{F_{1}, M_{1}, F_{2}, M_{2} \right\}
\label{eq:epsilon}
\end{equation}
In our ODE numerical integrations, we set the stopping condition at
$\epsilon=10^{-8}$.

Eqs.~(\ref{eq:model4}) assume that each individual encounters any
potential mate at the same average rate.  But full mixing will
seldom prove realistic, since mating encounters ordinarily occur
more frequently between nearby, than between distant pairs.
Spatially structured mating can be especially important during
ecological invasion, because introduced invaders often cluster
locally
\cite{GLO_JTB1999,KC_JTB2005,OMS_OIKOS2005,OBYKAC_TPB2006,ACK_EER2007}.
To address spatial detail, we generalized Eqs.~(\ref{eq:model4}) as
a reaction-diffusion system \cite{Murray_2003}. To model spatially structured mating
encounters, we replaced the homogeneous global densities with the
corresponding local densities ($F_{i}({\bf x})$, $M_{i}({\bf x})$)
at location ${\bf x}$.  To model dispersal we added a diffusion term
($D_{\rm diff}\nabla^2F_{i}({\bf x})$ and $D_{\rm diff}\nabla^2M_{i}({\bf x})$ for group $i$) to the
respective equation of motion.  To integrate the spatial model
numerically, we discretized the partial differential equations
(PDEs) to ODE equations (based on the Method of Lines technique
\cite{method_of_lines}) on a rectangular grid of size $400 \times
400$ (representing an area of $100 \times100$ units), using Neumann
boundary conditions. We integrated the resulting ODEs using an
explicit Euler time stepping, for which we chose a sufficiently
small time step ($\Delta t = 0.01$). These parameters allow us to
use diffusion coefficients as large as $2.5$ without producing
finite-size effects, or instability. For the spatial model, we
defined global equilibria with the stopping condition
$\epsilon=10^{-6}$.

\section*{Results} %%%%%%%%%%%%%%%%%%%%%%%%%%%%%%%%%%%%%%%%%%%%%%%%%%%%%%%
\subsection*{Stability of the resident}

\begin{figure}[t]
\begin{center}
\includegraphics[width=4in]{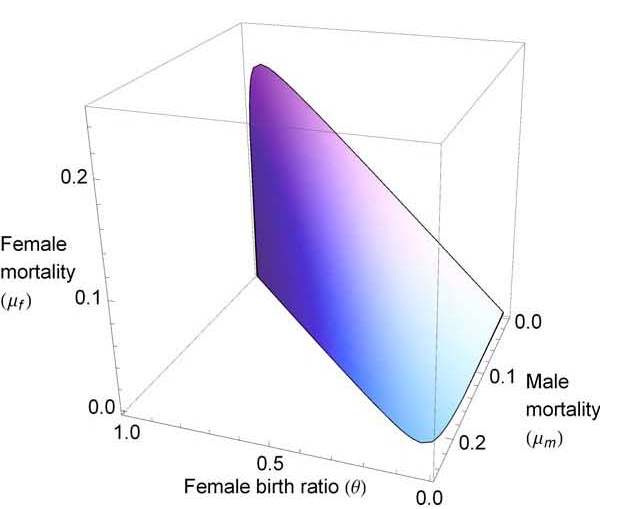}
\end{center}
\caption{{\bf Region of the parameter space where the resident is persistent.}
Parameter space region defined by Expression
(\ref{eq:real-cond}). Choosing parameters from the indicated domain
always results in a stable nonzero population, given sufficiently
high initial densities.}
\label{fig-paramspace}
\end{figure}

Before we address the dynamics of competitive invasion, we must
review \cite{Tainaka_EPL2006} and establish conditions for an
ecologically stable resident population. A stable resident occupies
the habitat alone, at a real, positive fixed point where
self-regulation limits growth, governed by Eqs.~(\ref{eq:model2}).
In general, the system has three fixed points: the trivial solution at zero density, and a pair of
nonzero fixed points. Extinction is always stable; one of the
nonzero fixed points is unstable, and the other one is stable. The
nonzero fixed points, hence a stable positive equilibrium, exist if (as shown in Supporting Information S1)
\begin{eqnarray}
D(\mu_{\rm f},\mu_{\rm m},\theta) =
1-4\left(\frac{\mu_{\rm f}}{\theta} + \frac{\mu_{\rm m}}{1 - \theta}\right) > 0 \;.
\label{eq:real-cond}
\end{eqnarray}
The necessary condition for this inequality is
\begin{equation}
\sqrt{\mu_{\rm f}} + \sqrt{\mu_{\rm m}} < 1/2 \;,
\end{equation}
in which case there exists a female-ratio continuum, $\theta_{\rm
c1}(\mu_{\rm f},\mu_{\rm m})<\theta<\theta_{\rm c2}(\mu_{\rm
f},\mu_{\rm m})$, where the population might persist. ``Might
persist'' means that a positive equilibrium exists, and initial
conditions determine whether or not the positive equilibrium
attracts the dynamics. If expression (\ref{eq:real-cond}) fails to
hold, the system exhibits only the trivial fixed point, stable
extinction. A resident population's persistence, then, depends on
interaction of the female ratio at birth with the sex-specific
mortality rates. In particular, when expression (\ref{eq:real-cond})
holds, any increase in the culturally transmitted mortality trait
$\mu_{\rm m}$ shrinks the range of female ratios maintaining an
extant resident population (see Supporting Information S1).
More generally, Figure \ref{fig-paramspace} depicts the region of the
parameter space satisfying expression (\ref{eq:real-cond}).
%%%%%%%%%%%%%%%%%%%%%%%%%%%%%%%%%%%%%%%%%%%%%%%%%%%%%%%
%%%%%%%%%%%%%%%%%%%%%%%%%%%%%%%%%%%%%%%%%%%%%%%%%%%%%%%%%

We performed a linear stability analysis of the system, using
Mathematica \cite{math}.  The results show that if condition
(\ref{eq:real-cond}) is met, then the larger (``$+$") roots in
Eqs.~(S6) (provided in Supporting Information S1)
are always locally stable, and the smaller roots are always unstable.

We present analytical formulae for the stable stationary densities
in Supporting Information S1. We used those formulae to quantify our numerical integration's
accuracy. We performed $5000$ test runs with randomly chosen
parameters that obey Expression (\ref{eq:real-cond}). For
$\epsilon=10^{-8}$, we find that the absolute difference of the
numerically computed fixed point was only $9.5\times10^{-7}\pm14\%$
from the analytical value, with $95\%$ confidence. This accuracy
suffices for our work.

To reach stable, positive equilibrium, population growth must
overcome a strong Allee effect \cite{BerecTREE_2006}, which defines
a separatrix on the phase map of initial female and male densities.
Below the separatrix extinction always results, independently of
other parameters, since growth is negative. Above the separatrix the
population grows to self-regulated equilibrium. To find this
threshold numerically, we select model parameters and fix the
initial female density. Then we conduct a binary search for the
initial male-density threshold value, numerically integrating Eqs.
(\ref{eq:model2}) until they converge to a stationary value (zero or
nonzero). Using this method we can determine the threshold value
with arbitrary precision.

\begin{figure}[t]
\begin{center}
\includegraphics[width=6.6in]{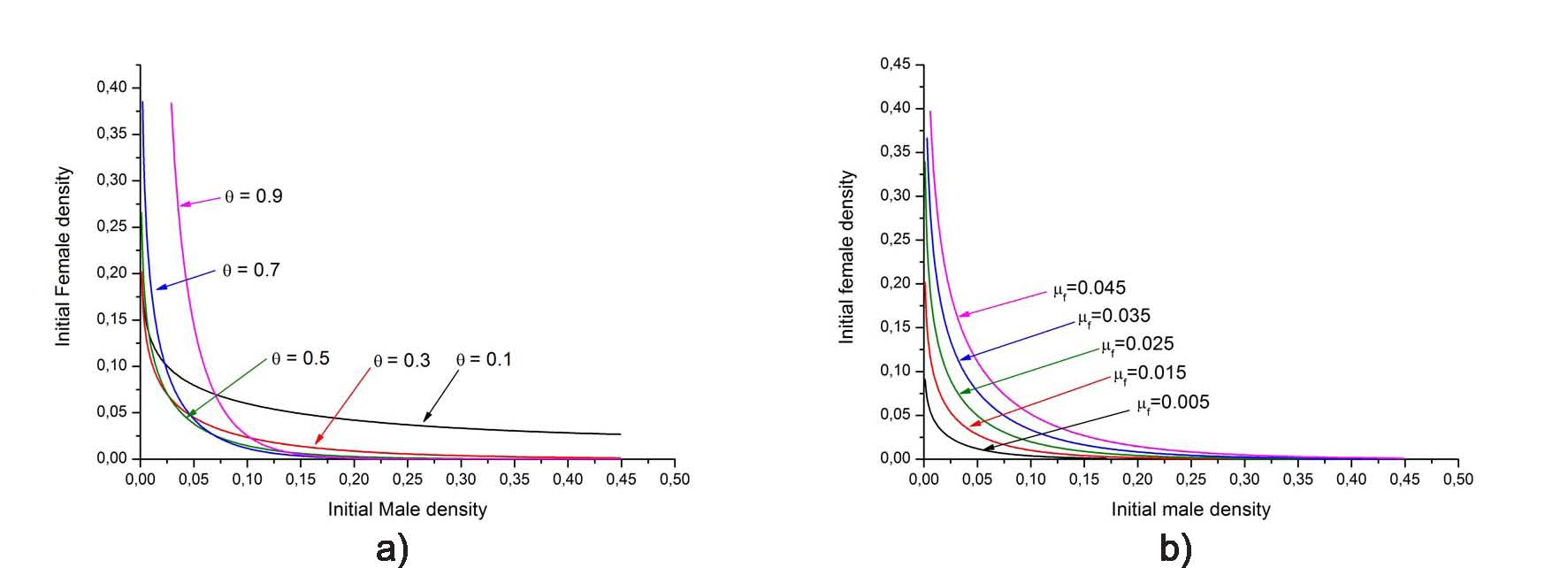}
\end{center}
\caption{{\bf Allee threshold of the resident.}
Survival/extinction threshold defined by the Allee-effect,
at various female ratios (a) and various
female mortalities (b). Other parameters:
(a): $\mu_{\rm f} = \mu_{\rm m} = 0.02$;
(b): $\theta=0.5$, $\mu_{\rm m} = 0.02$.}
\label{fig:allee}
\end{figure}

Figure \ref{fig:allee} displays the Allee-threshold for various
parameter combinations. In Fig. \ref{fig:allee}(a), where $\mu_{\rm
f} = \mu_{\rm m}$, an unbiased female ratio $(\theta = 0.5)$ allows
the lowest total population density before extinction due to the
Allee effect ensues.  When the sexes have the same mortality,
unbiased sex allocation also maximizes total population density at
positive equilibrium \cite{Tainaka_EPL2006}.

Figure~\ref{fig:allee}(b) verifies that increasing female mortality,
$\mu_{\rm f}$, for given $\theta$ and $\mu_{\rm m}$, expands the
region where the Allee effect leads to extinction.  Not
surprisingly, increasing male mortality produces a parallel effect.
Mortality-rate asymmetry and biased female ratios distort the shape
of the thresholds in Fig.~\ref{fig:allee}, but the same general
patterns emerge.

For a resident population, we have specified how existence of a
positive equilibrium depends on the interaction of female ratio at
birth and sex-specific mortalities.  We also have shown that initial
conditions (given existence of a positive equilibrium) required to
avert extinction due to the Allee effect depend on the same
parameters.  A practical consequence is that we must choose initial
densities for numerical integration carefully, so that when the
competitive dynamics results in extinction, we can clearly identify
the reason as either the Allee effect or exclusion.
%%%%%%%%%%%%%%%%%%%%%%%%%%%%%%%%%%%%%%%%%%%%%%%%%%%%%%%%%%%%
%%%%%%%%%%%%%%%%%%%%%%%%%%%%%%%%%%%%%%%%%%%%%%%%%%%%%%%%%%%%%%

\subsection*{Ecological competition: female ratio and invasion} %%%%%%%%%%%%%%%%%%%%%%%%%%%%%%%%%%%%%%%%%%

To quantify how population consequences of female-ratio evolution
can be affected by male mortality, we must have an ecological
understanding of the two-group competition model,
[Eqs.~(\ref{eq:model4})].  The system has nine fixed points; see
Supporting Information S1.  One is the trivial fixed
point where all densities vanish.  We can easily identify four more
fixed points related to those of the single-class case; there are
two symmetric pairs.  At these fixed points, competitive exclusion
leaves one group extinct, and one extant.  Exclusion implies that
one group's female ratio allele and its male-mortality cultural
trait have both gone extinct. Only one of these four, non-trivial
fixed points is locally stable: the ``$+$" solution
[Eq.~(S11)] of the group with the {\em greater  female
ratio}.  Assuming that $\theta_2 > \theta_1$, a necessary condition for this fixed point's local
stability is $\mu_2/\mu_1<2$ (see Supporting
Information for details). We shall refer to a
fixed point where one allele/culture persists after excluding the
other as a type-I fixed point.  When male mortality rates imply a
type-I fixed point, the greater female ratio always excludes the
lesser ratio.

\begin{figure}[t]
\begin{center}
\includegraphics[width=3.2in]{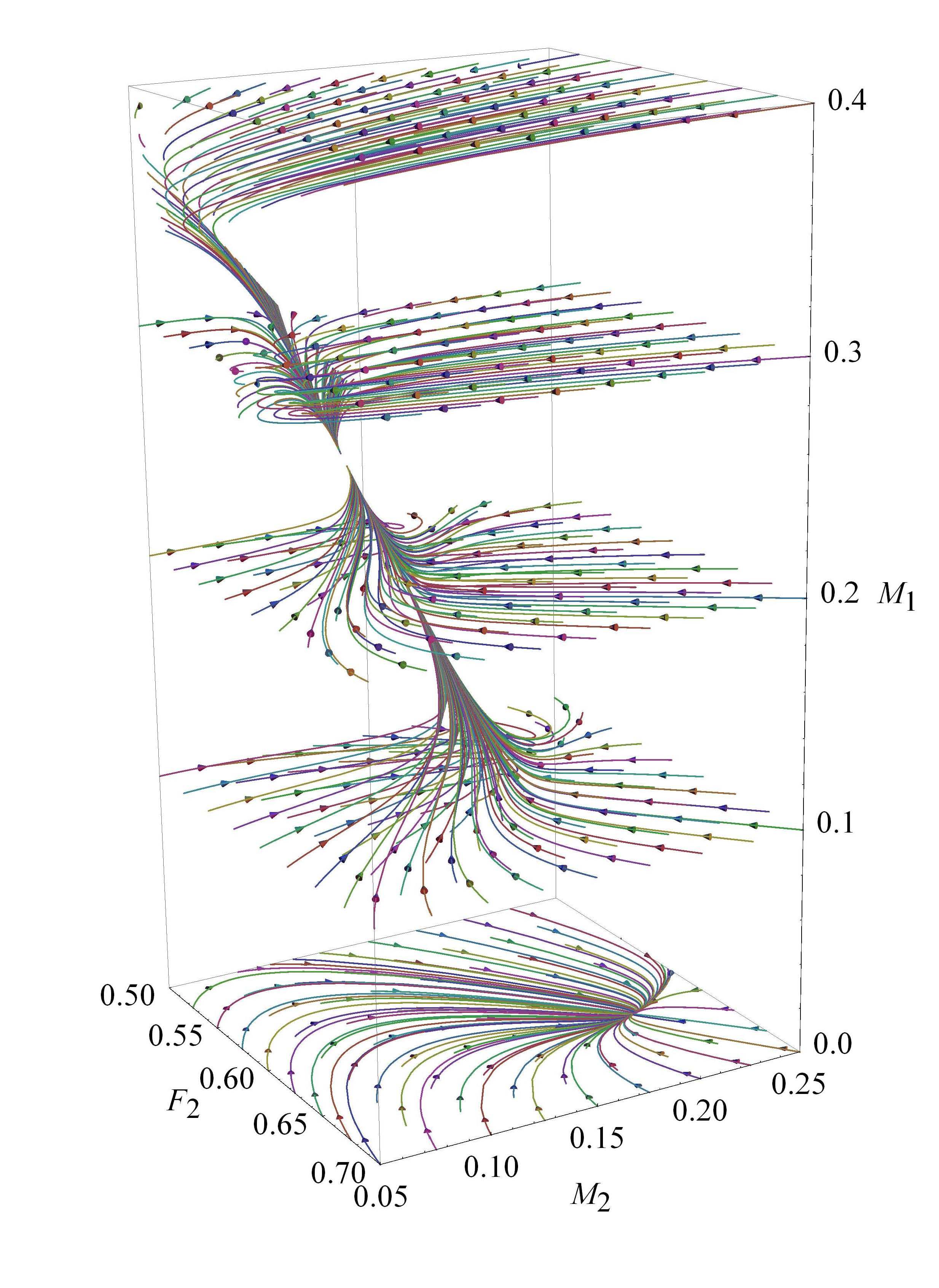}
\end{center}
\caption{{\bf Mean-field density flows.}
Density flows in the $(M_1,F_2,M_2)$ space (restricted to $F_1\equiv0$)
with type-I (``saddle", $M_1=0$) and type-II (``stable", $M_1>0$) fixed points for
$\mu_{\rm f}=0.02$, $\mu_1=0.01$, $\mu_2=0.04$;
$\theta_1=0.4$, $\theta_2=0.6$.}
\label{fig:flow}
\end{figure}

The four remaining fixed points (again, forming two pairs by
symmetry) are qualitatively distinct from those discussed above.  At
these fixed points only one female-ratio allele remains extant, but
male mortality traits ``coexist.''  That is, the population is
genetically uniform, in that all females carry the same female ratio
allele.  But the (male) population is culturally dimorphic; father
to son transmission [see Eqs.~(\ref{eq:model4})] maintains the
culture of the group whose females have been excluded competitively.
Consider a stable fixed point of this sort, when $F_1=0$.  The
necessary conditions are $\mu_2/\mu_1>2$, $\theta_2>\theta_1$, and
positivity of the discriminant
\begin{equation}
\tilde{D}(\mu_{\rm f},\mu_1,\theta_2) = %D(\mu_{\rm f},2\mu_1,\theta_2) =
1-4\left(\frac{\mu_{\rm f}}{\theta_2} + \frac{2\mu_1}{1-\theta_2} \right) >0 \;.
\label{coex_cond}
\end{equation}
The preceding condition holds if
\begin{equation}
\sqrt{\mu_{\rm f}}+\sqrt{2\mu_{1}}<1/2
\end{equation}
and $\tilde{\theta}_{\rm c1}(\mu_{\rm f},\mu_1)<\theta_2<\tilde{\theta}_{\rm c2}(\mu_{\rm f},\mu_1)$.
For mathematical details, see Supporting Information S1.  We refer to
stable fixed points combining a single female ratio and a male
cultural dimorphism as type-II fixed points.  Summarily, the model
does not permit equilibrium coexistence of female ratio alleles, but
can permit equilibrium diversity in cultural traits governing male
mortality. Also note, as is clear from the above conditions, that of
type-I and type-II fixed points {\em only one} can be stable at a
time. In Fig.~\ref{fig:flow} we illustrate the flow in the mean-field
dynamics for a set of parameters when both type-I and type-II fixed
points exits, but in the presence of co-occurring males of the other
allele, only type-II is stable.
%%%%%%%%%%%%%%%%%%%%%%%%%%%%%%%%%%%%%%%%%%%%%%%%%%%%%%%%%%%%
%%%%%%%%%%%%%%%%%%%%%%%%%%%%%%%%%%%%%%%%%%%%%%%%%%%%%%%%%%%%%%

Having obtained the nine fixed points for the two-group model
analytically, we approached the stability analysis numerically.
Analytical study of the system's stability proves difficult, due to
the number of variables and parameters (4 variables and 5
parameters). To be as thorough as possible, we performed numerical
integration systematically to span a significant region of the
five-dimensional parameter space. The range and step of the
parameters in our numerical scheme can be found in Table
\ref{tab:param-range}.
%%%%%%%%%%%%%%%%%%%%%%%%%%%%%%%%%%%%%%%%%%%%%%%%%%%%%%%%%%%%%%%%
\begin{table}[t]
\center
\small
\begin{tabular}{| c | c | c | c |}
\hline
Parameter & Lower bound & Upper bound & Step \\
\hline
\multicolumn{4}{|c|}{Series 1} \\
\hline
$\theta_{1}$ & $0.01 $ & $0.99$ & $0.01$ \\
\hline
$\theta_{2}$ & $0.01 $ & $0.99$ & $0.01$ \\
\hline
$\mu_{1}$ & $0.01 $ & $0.04$ & $0.005$ \\
\hline
$\mu_{2}$ & $0.01 $ & $0.04$ & $0.005$ \\
\hline
$\mu_{f}$ & $0.01 $ & $0.04$ & $0.005$ \\
\hline
\multicolumn{4}{|c|}{Series 2} \\
\hline
$\theta_{1}$ & $0.1 $ & $0.9$ & $0.1$ \\
\hline
$\theta_{2}$ & $0.1 $ & $0.9$ & $0.1$ \\
\hline
$\mu_{1}$ & $0.001 $ & $0.1$ & $0.001$ \\
\hline
$\mu_{2}$ & $0.001 $ & $0.1$ & $0.001$ \\
\hline
$\mu_{f}$ & $0.01 $ & $0.04$ & $0.01$ \\
\hline
\end{tabular}
\caption{{\bf Parameter regions and step sizes for numerical integration.}
Each set of parameters identifies two runs: one with high ($0.45$) and one with low ($10^{-4}$)
initial invader density.}
\label{tab:param-range}
\end{table}
%%%%%%%%%%%%%%%%%%%%%%%%%%%%%%%%%%%%%%%%%%%%%%%%%%%%%%%%%%%%%%%%%%%%
Each run begins with a stationary resident population, with allele
$1$ and cultural trait $\mu_1$.  If model parameters allowed a
stable positive equilibrium, we chose initial densities accordingly.
We then introduce the invaders, with female-ratio allele $2$ and
cultural trait $\mu_2$.  For each set of parameters (in each series)
we performed two runs, one with infinitesimal initial density of
invaders ($10^{-4}$) and one with high invader density ($0.45$).

\begin{figure}[t]
%\begin{center}
\hspace*{2cm}\includegraphics[width=4in]{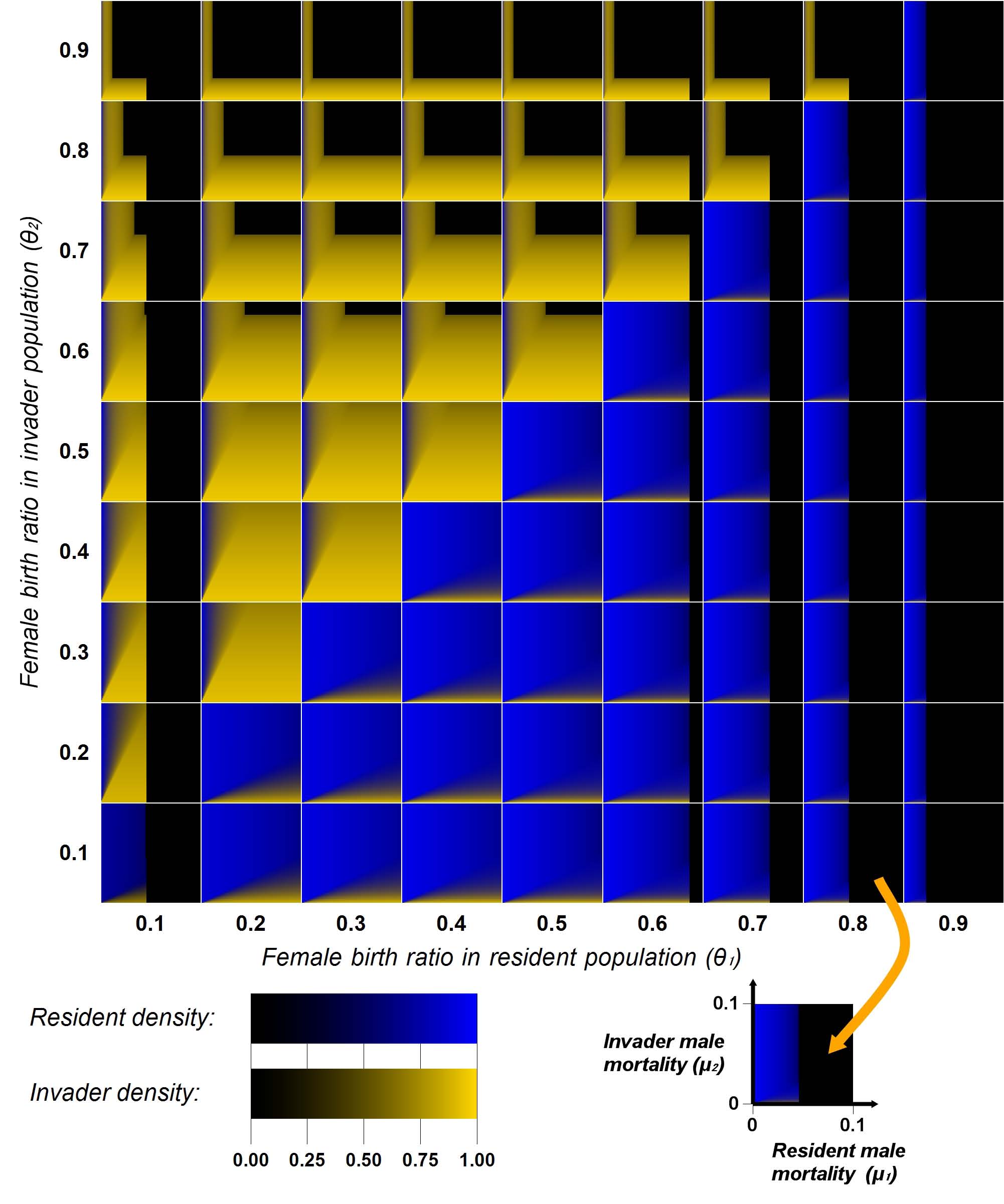}
%\end{center}
\caption{{\bf Stationary population densities.} Numerical
integrations are performed for the scenario where the persistent
stationary resident (group 1) is invaded by group 2, initially at an
infinitesimal density ($10^{-4}$). Large axes indicate common
parameters in rows and columns; every tile has the same axes, scaled
as indicated in the bottom right corner. Color scales use
independent color channels, therefore, resident and invader
densities are shown independently. Female mortality is fixed:
$\mu_{f}=0.02$.}
\label{fig:map}
\end{figure}

To portray the results, we generated a number of ``4D'' plots.  Each
shows a table containing 2D plots with the results of each run; the
axes of each 2D plot are values of the same two cultural parameters
($\mu_1$ and $\mu_2$, all with the same range).  Another two
parameters (female ratios $\theta_1$ and $\theta_2$) vary across the
rows and columns of the tables (the 4D plots).  We produced as many
tables as required by the range of the fifth parameter (female
mortality $\mu_{\rm f}$).  In each 2D plot, one pixel represents the
final stationary densities of the female ratio alleles.  The pixel's
location corresponds to the parameters for which it was computed;
resident and invader allele densities are shown on different color
channels. This way, we can visually compare all the results
simultaneously, simplifying the analysis greatly. Fig.~\ref{fig:map}
shows one 4D plot; the associated parameter ranges produce the full
set of the model's outcomes.
%%%%%%%%%%%%%%%%%%%%%%%%%%%%%%%%%%%%%%%%%%%%%%%%%%%%%%%%%%%%%%%%%%%%%
%%%%%%%%%%%%%%%%%%%%%%%%%%%%%%%%%%%%%%%%%%%%%%%%%%%%%%%%%%%%%%%%%%%%

In what follows, we investigate the necessary and sufficient
conditions for successful (pairwise) genetic invasion of the
resident female ratio, and the necessary conditions for cultural
``coexistence."

\subsubsection*{Invasion and exclusion} %%%%%%%%%%%%%%%%%%%%%%%%%%%%%%%%%%%%%%%%%

Our numerical results reveal immediately that female ratios
determine the outcome of invasion; a successful invader in pairwise
competition has the greater female ratio.  That is, successful
invasion always requires $\theta_2$$>$$\theta_1$, and
$\theta_2$$<$$\theta_1$ assures that the resident resists invasion.
When the invader has the greater female ratio, it excludes the
resident allele competitively.  Furthermore, successful invasion by
a female-ratio allele assures that the associated cultural trait
(with value $\mu_2$) advances from rarity.  As a numerical check, we
note that both infinitesimal and high invader densities always
result in identical final densities.

Since the female-ratio allele is sex-linked, dependence of invasion
on $(\theta_2 - \theta_1)$ simply recalls Hamilton
\cite{Hamilton_1967}.  But in our model, the ecological effect of
invasion depends on the culturally transmitted trait.  Suppose that
successful invasion excludes both the resident female ratio allele
($\theta_2$$>$$\theta_1$) and the resident cultural trait ($F_1$$=$$0$, $M_1$$=$$0$).
From Supporting Information S1, the necessary
conditions for invasion and combined genetic/cultural exclusion (type-I fixed point) are:
\begin{equation}
\sqrt{\mu_{\rm f}}+\sqrt{\mu_{2}}<1/2 , \quad\quad\quad\quad
\mu_{2}/\mu_{1} < 2 , \quad\quad \mbox{and} \quad\quad
\theta_2>\theta_1 \;.
\label{invasion_conditions}
\end{equation}
Sufficient conditions for invasion and exclusion of both resident
traits further require: $\theta_{\rm c1}(\mu_{\rm
f},\mu_2)<\theta_2<\theta_{\rm c2}(\mu_{\rm f},\mu_2)$, ensuring
that the invader attains positive stable equilibrium.

\begin{figure}[t]
%\begin{center}
\hspace*{1cm}\includegraphics[width=4.9in]{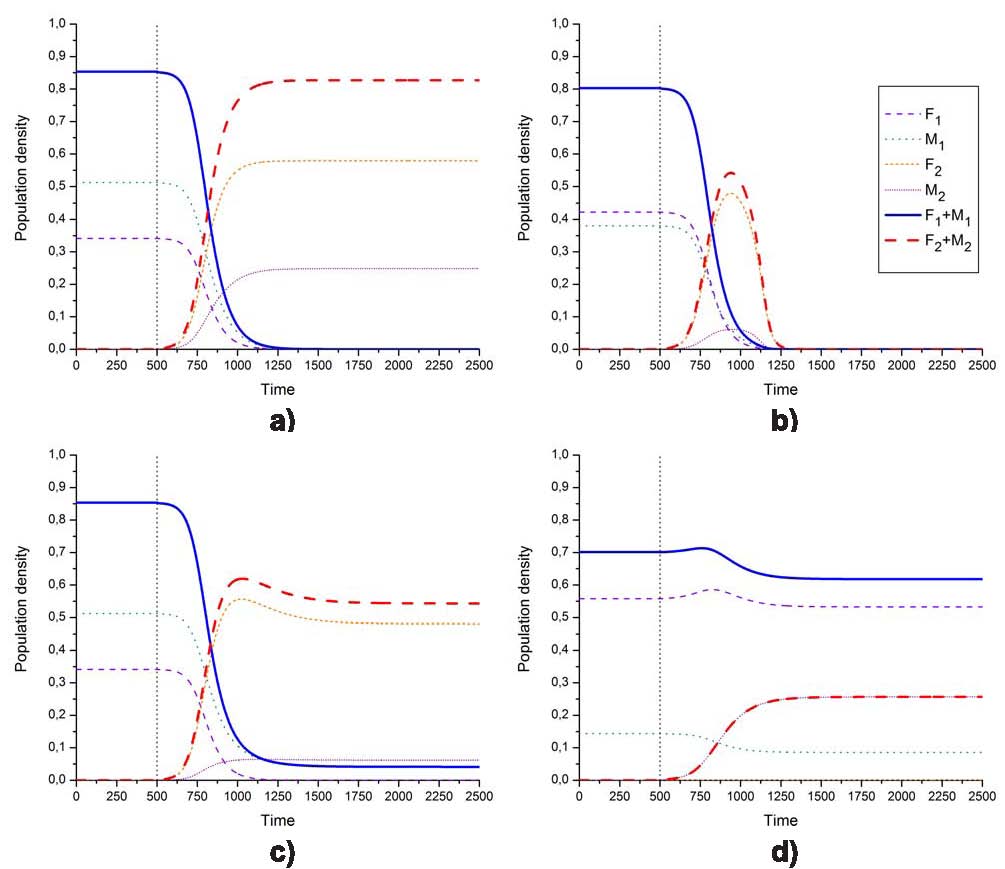}
%\end{center}
\caption{{\bf Population-density time series.}
Panel (a) shows successful
invasion; (b) shows invasion followed by extinction; (c) shows
coexistence of resident males with the invader allele; (d) shows
coexistence of invader males with resident allele. The vertical
dotted line indicates the time when the invader was added to the
system, at $10^{-3}$ density (both males and females). Legends shown
to the right of panel (b) describe data on all four panels.
Common parameter: $\mu_{f} = 0.03$. Individual parameters:
(a) $\theta_{1} = 0.4$, $\theta_{2} = 0.7$, $\mu_{1} = 0.03$, $\mu_{2} = 0.03$;
(b) $\theta_{1} = 0.4$, $\theta_{2} = 0.7$, $\mu_{1} = 0.05$, $\mu_{2} = 0.08$;
(c) $\theta_{1} = 0.4$, $\theta_{2} = 0.7$, $\mu_{1} = 0.03$, $\mu_{2} = 0.08$ (here, $F_1$$=$$0$ in the final equilibrium);
(d) $\theta_{1} = 0.7$, $\theta_{2} = 0.4$, $\mu_{1} = 0.05$, $\mu_{2} = 0.01$ (here, $F_2$$=$$0$ in the final equilibrium).}
\label{fig:timeseries}
\end{figure}

Figure \ref{fig:timeseries}(a) shows an example of successful
invasion leading to exclusion of both the resident allele and
resident culture.  Following introduction of the invading group, the
resident density drops quickly, and the successful allele (females)
and successful culture (observed in males) advance to become the new
resident group.  Invasion, full exclusion of the resident, and
population persistence first require that the successful invader's
male mortality assures, given female mortality $\mu_{\rm f}$,
feasibility of a stable, positive equilibrium in the absence of
between-group competition.  Expression (S7) gives the explicit
cultural constraint on female ratios guaranteeing a stable, positive
equilibrium.  Assuming this condition holds, the invader must,
secondly, have the greater female ratio.  But the invader's
demographic advantage of a greater female ratio will not exclude
both the resident allele and resident culture unless constraints on
the mortality rate are satisfied.  Specifically, the invader's
cultural trait $\mu_2$ cannot exceed either $(\frac{1}{2} -
\sqrt{\mu_{\rm f}})^2$ nor $2\mu_1$.
%%%%%%%%%%%%%%%%%%%%%%%%%%%%%%%%%%%%%%%%%%%%%%%%%%%%%%%%%%%%%%%%%%%%%%%
%%%%%%%%%%%%%%%%%%%%%%%%%%%%%%%%%%%%%%%%%%%%%%%%%%%%%%%%%%%%%%%%%%%%%%%%%%%%

\subsubsection*{Invasion and cultural coexistence} %%%%%%%%%%%%%%%%%%%%%%%%%%%%%%%%%%%%%%%%

Recall that Eqs.~(\ref{eq:model4}) do not have fixed points where
differing female ratios co-occur. The model, however, does allow for
cultural coexistence, where males of both groups co-occur, but
females of only one group remain extant.  For details, see
Supporting Information S1.

In one such scenario, resident females are excluded ($F_1$$=$$0$),
but resident males, a cultural designation, persist ($M_1$$>$$0$). Necessary
conditions for this type of coexistence (\emph{i.e}., for a type-II
stable fixed point) are
\begin{equation}
\sqrt{\mu_{\rm f}}+\sqrt{2\mu_{1}}<1/2 , \quad\quad\quad\quad
\mu_{2}/\mu_{1} > 2 , \quad\quad \mbox{and} \quad\quad
\theta_2>\theta_1 \;.
\label{eq:coexistence_f1null}
\end{equation}
For sufficiency, the invaders' female ratio must fall into a finite
interval, $\tilde{\theta}_{\rm c1}(\mu_{\rm
f},\mu_1)<\theta_2<\tilde{\theta}_{\rm c2}(\mu_{\rm f},\mu_1)$,
given by the positivity requirement of the corresponding
discriminant [Eq.~(\ref{coex_cond})].

Figure~\ref{fig:timeseries}(c) displays an example where the
resident culture, but not the resident allele, persists after
successful invasion.  The invader has the greater female ratio, and
excludes the resident allele competitively.  The final equilibrium
state is a type-II fixed point where the resident's male-mortality
trait persists via father-to-son cultural transmission.  The ratio
of males at dynamic equilibrium is $M_{1}/M_{2}=\mu_{2}/\mu_{1} -
2$.  Note that the competitively driven increase in female ratio
produces a decrease in total population density (females plus males)
at equilibrium [Fig.~\ref{fig:timeseries}(c)].

By the symmetry of the equations, there also exists a type-II stable
fixed point with $F_2$$=$$0$. That is, the resident population
resists invading females, but the introduced cultural trait advances
from rarity.  Put simply, we can exchange the resident-invader roles
of the two groups, and reach the same dynamic equilibrium.
Necessary conditions for this case are
\begin{equation}
\sqrt{\mu_{\rm f}}+\sqrt{2\mu_{2}}<1/2, \quad\quad\quad\quad
\mu_{2}/\mu_{1} < 1/2, \quad\quad \mbox{and} \quad\quad
\theta_2<\theta_1  \;.
\label{eq:coexistence_f2null}
\end{equation}
Here, the introduced female ratio $(\theta_2 < \theta_1)$ is
repelled.  However, the invading male mortality culture, introduced
at infinitesimal density, advances and persists at equilibrium; see
Fig.~\ref{fig:timeseries}(d).  The ratio of males at this
equilibrium $M_{2}/M_{1}=\mu_{1}/\mu_{2} - 2$.

Figure \ref{fig:map} includes cases of equilibrium cultural
coexistence. For example, condition (\ref{eq:coexistence_f1null}) is
visible in tiles where $\theta_{1}=0.2$ and $\theta_{2}=0.4$;  the
sharp change in color along the line $\mu_{2} = 2 \mu_{1}$ indicates
the condition for cultural coexistence. When this condition is not
satisfied, the culture associated with the lower female birth ratio
always declines to extinction. In both cases, the fixed points found
numerically are identical to the analytical fixed points for the
respective equilibria: Eqs.~(S16) for cultural
coexistence, and Eq.~(S11) for competitive exclusion of
both allele and culture.

\subsubsection*{Invasion to extinction} %%%%%%%%%%%%%%%%%%%%%%%%%%%%%%%%%%%%%%%%%%%

Given the competitive advantage of increased female allocation in
our model, evolution of the sex-linked trait might threaten
population persistence.  Our model's dynamics includes a case where
successful invasion of a stable resident is followed by extinction
of the entire population. We observe this result in numerical
experiments where the invader has both the greater female ratio and
the greater male mortality rate, so that Expression
(\ref{eq:real-cond}) fails to hold.  The greater female ratio drives
invasion, but the invader's combined  genetic-cultural demography
does not satisfy the condition for a stable, positive equilibrium.
Hence, the successful invader would not advance from rarity absent
the resident group.

Figure~\ref{fig:map} shows an example of invasion to extinction;
note the black region of the tile where $\theta_{1}=0.4$ and
$\theta_{2}=0.7$.  For a particular mortality-rate combination,
Fig.~\ref{fig:timeseries}(b) depicts the time-dependent densities
for a case of invasion to extinction. The necessary conditions for
invasion, see Eq.~(\ref{invasion_conditions}), are met.  However,
$\theta_2 >\theta_{\rm c2}$.  Hence the invader grows when rare and
excludes the resident, but the invader cannot persist. Essentially,
the invading female ratio allele increases its initial density by
``exploiting'' males of the resident group while competing for
resources with resident females. After some time the density of the
resident females reaches zero. The reduced density of females means
that the production of males (both resident and invader) is reduced.
Consequently, the invading group, once occupying the environment
alone, cannot maintain a positive equilibrium density, and a
``marriage squeeze'' takes the population to extinction.

Given this result, one can envision a stable population where
immigration or mutation introduces new alleles over a lengthy time
scale.  If a new allele has a higher female ratio than the current
resident, it will advance. A series of allelic substitutions might
increase the female ratio continuously. Our model does not prevent
the female ratio from surpassing the threshold defined by
Eqs.~(\ref{eq:real-cond}), where the population begins to decline to
extinction --- recalling Hamilton's \cite{Hamilton_1967} comment on
sex linkage and sex-ratio evolution.
%Therefore, all populations are destined to go extinct, eventually.

\subsection*{Local mate density and spatial invasion} %%%%%%%%%%%%%%%%%%%%%%%%%%%%
\label{sec:spatial}

\subsubsection*{Invading an open habitat: the critical radius}

Equations (\ref{eq:model4}) and (\ref{eq:model2}) assume that
densities mix homogeneously, a strong simplification for most
organisms.  Furthermore, invasion most often has a distinctly
spatial character, expanding from one or more foci of introduction
\cite{OBYKAC_TPB2006}.  To consider both effects, we assumed a
two-dimensional habitat with local mating and random mobility of
individuals.  This elaborates our model as a reaction-diffusion
system \cite{Murray_2003}.
%%%%%%%%%%%%%%%%%%%%%%%%%%%%%%%%%%%%%%%%%%%%%%%%%%%%%%%%%%%%%%%%%%%%%%%%%%%%%%%%%%
Note, however, that our spatial but deterministic reaction-diffusion
equations still maintain an essential (local) ``mean-field
character" (in the statistical physics sense and terminology) in
that all correlation functions are still factorized into products of
concentrations \cite{McKane_PRE2004,{Korniss_PRE1997}}. A
stochastic, spatial individual-based model or its Langevin-type,
stochastic reaction-diffusion analogue (not addressed in this work)
may, in principle, lead to different behaviors
\cite{vKampen_1981,Hinrichsen_2000}. For example, the region of
persistence in the case of a single-group two-sex population becomes
significantly narrower in a stochastic lattice-based model
\cite{Tainaka_EPL2006}.
%%%%%%%%%%%%%%%%%%%%%%%%%%%%%%%%%%%%%%%%%%%%%%%%%%%%%%%%%%%%%%%%%%%%%%%%%%%%%%%%%%%

Successful invasion in spatial environments ordinarily requires that
an initial invader cluster have some minimal size for further growth
\cite{LK_TPB1993,GLO_JTB1999,OBYKAC_TPB2006,ACK_EER2007}.  This
criterion may be due to an Allee effect \cite{LK_TPB1993} or
inherent geometrical constraints on cluster expansion
\cite{ACK_EER2007}.  For systems exhibiting the Allee effect under
homogeneous mixing, one can specify this minimal cluster size as the
critical radius ($R_{\rm c}$) required for spatial invasion.
Assuming radially symmetric growth, one expects $R_{\rm c} \sim
\sqrt{D_{\rm diff}}$, where $D_{\rm diff}$ is the diffusion
coefficient \cite{LK_TPB1993}.  For simplicity, we take $D_{\rm
diff}$ as a constant across all individuals.  The first goal of our
spatial analysis was to confirm this scaling relationship for the
critical radius when a single group is introduced in an open
(unoccupied) habitat.

For spatial invasion in an open habitat, individuals diffusing away
from the perimeter of the invader cluster encounter mate densities
too low for population increase, given the Allee effect
(\emph{i.e}., extinction is stable).  A small invader cluster can
shrink as a result.  A cluster size exceeding the critical radius
generates interior densities sufficient to drive cluster expansion.
The critical radius depends on both density inside the cluster and
the diffusion coefficient.  Therefore, calculating a critical radius
demands specifying initial densities within the circular cluster.
We noted that as we chose densities closer to, but exceeding, the
Allee threshold of the homogenous-mixing case, the critical radius
increased. Therefore, a reasonable (deterministic) choice is the
stationary density of the non-spatial model, which we can calculate,
given the female ratio and sex-specific mortality rates [see
Eq.~(S6)].

\begin{figure}[t]
\begin{center}
\includegraphics[width=5in]{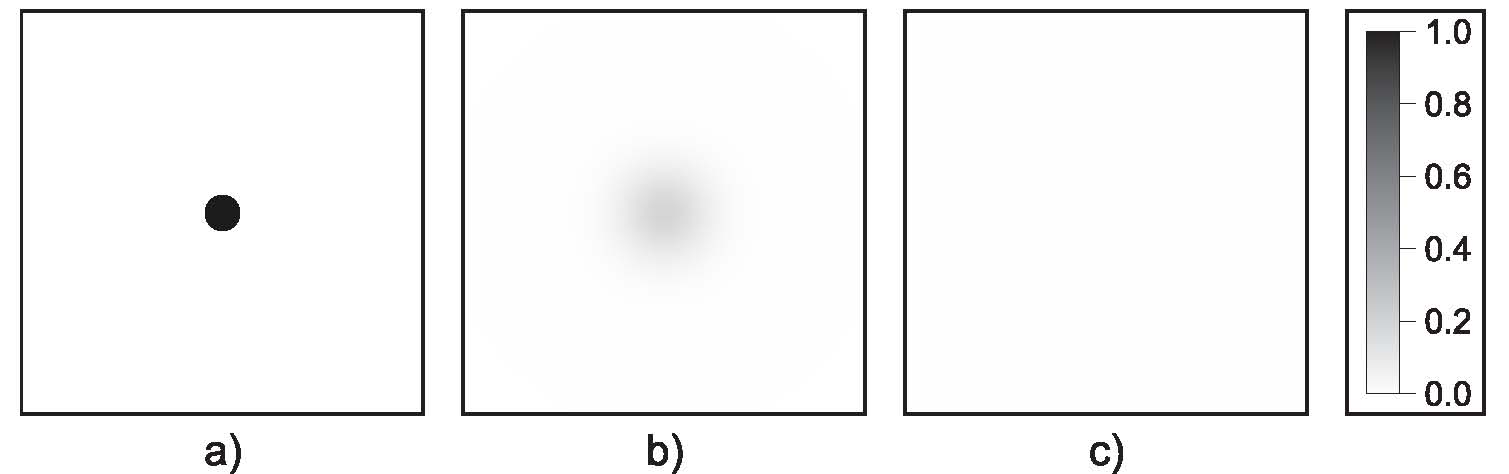}
\end{center}
\caption{{\bf Unsuccessful spatial persistence in the single-group system.}
Population dynamics in the single-group
system (open habitat), where the initial radius is less than the critical radius
($R_{0} = 4.5 < R_{\rm c} = 5.1$). Simulation time: (a) $t = 100$,
(b) $t = 150$, (c) $t = 300$. Parameters: $\theta = 0.5$, $\mu_{f} =
0.02$, $\mu_{m} = 0.03$, $D_{\rm diff} = 1.0$.}
\label{fig:DiffBelow}
\end{figure}

\begin{figure}[t]
\begin{center}
\includegraphics[width=5in]{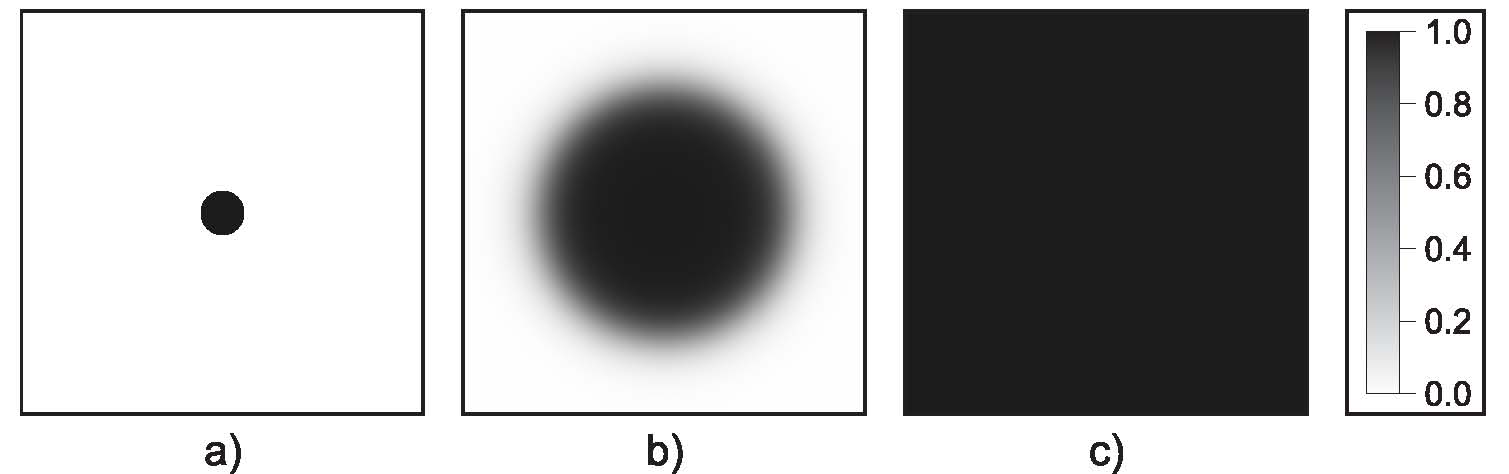}
\end{center}
\caption{{\bf Successful spatial persistence in the single-group system.}
Population dynamics in the single-group
system (open habitat), where the initial radius is greater than the critical radius
($R_{0} = 5.5
> R_{\rm c} = 5.1$). Simulation time: (a) $t = 100$, (b) $t = 300$,
(c) $t = 600$. Parameters: $\theta = 0.5$, $\mu_{f} = 0.02$,
$\mu_{m} = 0.03$, $D_{\rm diff} = 1.0$.}
\label{fig:DiffAbove}
\end{figure}

We found the critical radius by performing a binary search, using
the initial interval of $R \in [1, 20]$.  At each step, a simulation
runs with a particular initial radius, until all densities at all
grid points come to a stationary state (where all time derivatives
are less than $\epsilon=10^{-6}$). In this final state either all
grid points have the positive, stationary densities of the
non-spatial model, or all have zero densities. The resolution of the
grid ($4$ cells/unit distance) and the discretization of a circle on
a rectangular grid allow us to measure non-integer radii. Time
evolution of a shrinking ($R$$<$$R_{\rm c}$) and a successfully
growing, invading  population ($R$$>$$R_{\rm c}$) are illustrated in
Figs.~\ref{fig:DiffBelow} and \ref{fig:DiffAbove}, respectively.
%%%%%%%%%%%%%%%%%%%%%%%%%%%%%%%%%%%%%%%%%%%%%%%%%%%%%%%%%%%%%%%%%%%
%%%%%%%%%%%%%%%%%%%%%%%%%%%%%%%%%%%%%%%%%%%%%%%%%%%%%%%%%%%%%%%%
%%%%%%%%%%%%%%%%%%%%%%%%%%%%%%%%%%%%%%%%%%%%%%%%%%%%%%%%%%%%%%%%%%%%

\begin{figure}[t]
\begin{center}
\includegraphics[width=4in]{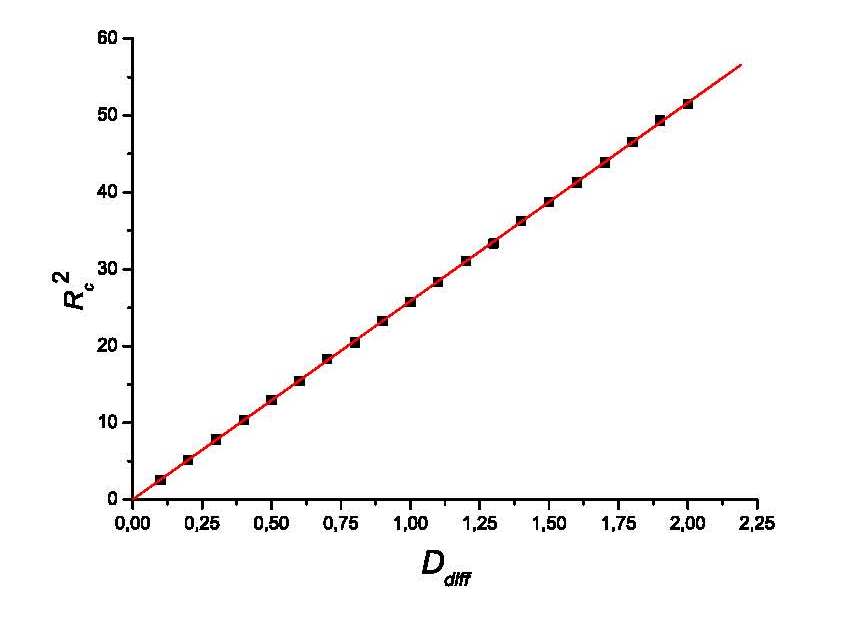}
\end{center}
\caption{{\bf Behavior of the critical radius of the single-group system.}
Square of the critical radius of a population at self-regulated equilibrium of a single-group system,
as a function of the diffusion coefficient (common for both sexes). The line shown is a fitting
using minimum least squares, with a correlation value of $0.99997$.
Parameters: $\theta = 0.5$, $\mu_{f} = 0.02$, $\mu_{m} = 0.03$.}
\label{fig:Rcrit}
\end{figure}

We obtained the critical radius for various diffusion coefficients,
at certain fixed set of parameters [Fig.~\ref{fig:Rcrit}]. As
anticipated \cite{LK_TPB1993}, the results confirm that the critical
radius is proportional to the square root of the diffusion
coefficient.
%%%%%%%%%%%%%%%%%%%%%%%%%%%%%%%%%%%%%%%%%%%%%%%%%%%%%%%%%%%%%%%%%%%%%
%%%%%%%%%%%%%%%%%%%%%%%%%%%%%%%%%%%%%%%%%%%%%%%%%%%%%%%%%%%%%%%%%%%%%

\subsubsection*{Spatial invasion of a resident population}

We extended the between-group competition model to the spatial case
with diffusion, using the same grid size, resolution, and diffusion
coefficients as we used in the open-habitat model. The goal here is
to ascertain if there is a critical radius for successful invasion
when invaders can mate with residents in an occupied habitat.

We initiated simulations differently than in the open-habitat case.
Here, every grid point was initialized to the stationary density of
the resident group.  Then, we introduced the invader within a circle
of a given radius, at a small density. The simulation ran until all
grid points come to a stationary state (where all time derivatives
are less than $10^{-6}$).

\begin{figure}[t]
\begin{center}
\includegraphics[width=5in]{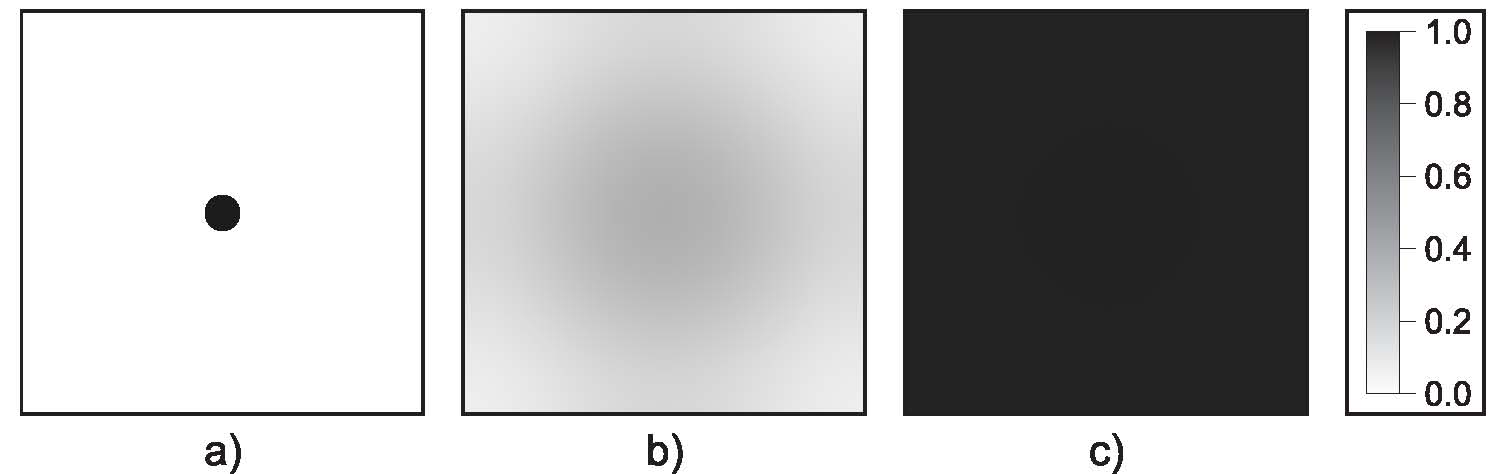}
\end{center}
\caption{{\bf Spatial invasion in the two-group system.}
Evolution of the invader population density while invading a stable resident
population. [Note that the initial radius is less than the critical
radius ($R_{0} = 4.5 < R = 5.1$) for invasion into an open habitat]. For clarity, only the invaders' density is shown.
Simulation time: (a) $t = 100$, (b) $t = 450$, (c) $t
= 820$. Parameters: $\theta_{1} = 0.3$, $\theta_{2} = 0.5$, $\mu_{f}
= 0.02$, $\mu_{1} = \mu_{2} = 0.03$, $D_{\rm diff} = 1.0$.}
\label{fig:DiffInvade}
\end{figure}

We found that no matter how small we set the invader density and
cluster radius, the result was always identical to the homogeneously
mixed case.  That is, the allele with the higher female ratio
persists, and the ecological impact of the winning female ratio
depends on the male mortality rates.  Male cultural traits may
coexist (type-II fixed point), or both females and males of
the lower female-ratio group go extinct (type-I fixed point).
Figure \ref{fig:DiffInvade} shows a scenario where the invader has
the same parameters as the open-habitat invasion in
Fig.~\ref{fig:DiffBelow}.  However, the result is different, because
of the presence of the resident population. The invader can
(effectively) exploit the resident population as mates, enabling the
invader to spread successfully and eventually exclude the resident.
%%%%%%%%%%%%%%%%%%%%%%%%%%%%%%%%%%%%%%%%%%%%%%%%%%%%%%%%%%%%%%%%%%%%
%%%%%%%%%%%%%%%%%%%%%%%%%%%%%%%%%%%%%%%%%%%%%%%%%%%%%%%%%%%%%%%%%%%%%%

We understand the absence of a critical radius in the
resident-occupied environment by considering cases where even an
infinitesimal invader density can completely exclude the resident in
the homogenously mixed case.  In the worst-case scenario (for the
invading allele and culture), we introduce only a small density of
invaders at only a single grid point, with a high diffusion rate.
Then, diffusion spreads the invader to all grid points, making its
density extremely small, but greater than zero. However, this is
enough for successful invasion at every grid point, independently of
other locations, as we noted in the model with global mixing. If we
introduce a greater density of invaders, with slower diffusion, then
the invader can quickly overtake the local area before spreading out
as a diffusive front.  The eventual result will be the same.  Hence
we conclude that there is no critical radius for invasion with
diffusion, if a resident population already occupies the habitat.

\section*{Discussion} %%%%%%%%%%%%%%%%%%%%%%%%%%%%%%%%%%%%%%%%%%%%%%%%%%%%%%%
Most models of sex ratio evolution, whether analyzed as
evolutionarily stable sex allocation \cite{Charnov_1982,West_2009}
or developed with population-genetic detail \cite{Karlin_1986},
assume that a parent is related symmetrically to female and male
offspring.  Hamilton \cite{Hamilton_1967} noted that sex-linked
inheritance of a gene for sex ratio breaks this symmetry, and
extraordinary sex ratios can evolve as a consequence.  Frank
\cite{Frank_1986}  summarizes effects of asymmetric relatedness to
offspring by sex, and cites several studies where this asymmetry is
correlated with strongly biased investment in the sexes; see
Uyenoyama and Bengtsson \cite{Uyenoyama_1982}.  Our results specify
how the degree of bias can interact with a between-sex mortality
difference to influence the population dynamic consequences of sex
ratio evolution.

Tainaka et al. \cite{Tainaka_EPL2006} and Nitta et al.
\cite{Nitta_LNCS2008} developed spatially detailed models to study how
sex ratio might affect population persistence.  For successful
mating, their model requires that at least one fertile individual of
each sex occupy a site neighboring an empty site (where the
offspring is placed).  At the scale of individuals, the dynamics is
the simplest generalization of the contact process
\cite{Harris_AP1974,Liggett_1999,MD_1999} that can capture both
two-sex reproduction and preemptive competition
\cite{OMS_OIKOS2005,KC_JTB2005,OBYKAC_TPB2006}. Given female and
male mortality rates, they find the sex ratio maximizing population
density, and note that sex ratios differing too much from this
singular value lead to population extinction \cite{Tainaka_EPL2006}.
Compared to the mean-field result, the extinction effect due to
biased sex ratio sharpens in simulation of the stochastic,
lattice-based model; the range of sex ratios producing population
persistence becomes quite narrow.  Since mating pairs form locally,
biasing the sex ratio rapidly diminishes the chance that an open
site will be neighbored by one individual of each sex.  So,
demographic stochasticity may lead to extinction once sex ratio is
biased, and genetic drift may permit biased sex ratios to evolve
even when bias is selectively disfavored \cite{Lande_1998}.

Our study generalizes the model of Tainaka et al.
\cite{Tainaka_EPL2006} by including between-sex differences in
mortality and detailing outcomes of competition between different
female ratios.  Our model limits expression of sex ratio to the
heterogametic sex, so that stronger bias in sex allocation has a
competitive advantage.  Our results elucidate the ecological effects
of interaction among the degree of sex ratio bias and sex-specific
mortality for competitive/cultural invasion and demographic
stability. In the simplest case, an introduced  female allocation
and associated cultural trait, male mortality, invades and excludes
the resident allele and culture.  Complete exclusion requires only
that the invaders have the higher female allocation  and that their
male mortality rate is lower than twice that of the resident males.
If the invader's male mortality rate is large enough to exceed this
limit, but the difference in female allocation remains, the resident
culture (but not the resident allele) survives and coexists with the
invader's culture.

Our analysis also identified an interesting invasion-to-extinction
scenario.  A group with the greater female allocation and greater
male mortality (compared to the demographically stable resident)
cannot invade an empty environment.  Yet it invades and excludes the
resident, and then goes extinct, because of its high female ratio.
Since the invaders can mate with the residents, they effectively
exploit the resident group in the early phase of invasion and, when
sufficiently numerous, drive the resident extinct.  Thereafter, a
marriage squeeze leaves the invader declining to extinction.  This
type of outcome, where sex ratio and an Allee effect can push a
population to extinction, may have application in the management of
pest populations \cite{BerecTREE_2006}.  Evolutionarily, the
demographic consequences of sex ratio bias may favor suppression of
sex-ratio distorters \cite{Hamilton_1967}, and may promote (or be
tolerated by) clonal reproduction \cite{Shelton_2010}.

%%%%%%%%%%%%%%%%%%%%%%%%%%%%%%%%%%%%%%%%%%%%%%%%%%%%%%%%%%%%%%%%%%%%%%%

The basic two-group two-sex model we considered in this work also
allows for some straightforward, yet rich generalizations. In this
paper we focused on the scenario where following mating between
females and males of different groups, male offspring acquire either
cultural trait with probability $1/2$. To capture asymmetry in the
biparental transmission of the cultural trait in males, our model
and the corresponding equations can be generalized to an asymmetric
case where male offspring resulting from mating between a female of
group $i$ and a male of group $j$ acquire the cultural trait of
group $i$ or group $j$ with probability $p$ and $q$, respectively
($p+q=1$). (Vertical cultural-transmission probabilities can,
indeed, vary across different combinations of parental phenotypes
\cite{LCS_1981}.) While we do not analyze this asymmetric model in
detail, we included the corresponding homogeneous mean-field
equations and their fixed points in Supporting Information S1 with
the basic findings and note that the qualitative behavior of the
system remains the same. In particular, both type-I and type-II
fixed points exist, corresponding to full invasion/exclusion and
partial invasion/cultural coexistence, respectively. Naturally, for
$p$$>$$q$ ($p$$<$$q$) the size of the parameter region with cultural
coexistence narrows (widens) and the size of the surviving and
coexisting resident culture decreases (increases).
%%%%%%%%%%%%%%%%%%%%%%%%%%%%%%%%%%%%%%%%%%%%%%%%%%%%%%%%%%%%%%%%%%%%%%

{\bf Note:} The above generalization (asymmetric cultural
transmission in cross-cultural mating) was suggested by an anonymous
referee during the review process of this paper.

\section*{Acknowledgments}

This research was supported by the National Science Foundation under
Grants Nos. DEB-0918413 and DEB-0918392.

\section*{Author contributions}

Conceived and designed models: TC GK. Performed numerical
experiments: FM GK. Analyzed and discussed numerical and analytic
results: FM TC GK. Wrote manuscript: FM TC GK.

\newpage
%%%%%%%%%%%%%%%%% SUPPLEMENTARY INFORMATION %%%%%%%%%%%%%%%%%%%%%%%%%%%%%%%%%%%%%%%%%%%%%%%%%%

%\appendix

%\renewcommand{\figurename}{Fig. S\hspace*{-0.10truecm}}\setcounter{figure}{0}

\renewcommand{\thefigure}{S\arabic{figure}}\setcounter{figure}{0}
\renewcommand{\theequation}{S\arabic{equation}}\setcounter{equation}{0}
% (so that equations are numbered (S1), (S2), ...

\renewcommand{\thesection}{S\arabic{section}}\setcounter{section}{0}

\setcounter{page}{1}

\begin{flushleft}
{\Large
\textbf{Supporting Information S1:\\
Analysis of the Mean-Field Fixed Points}
\\
\vspace*{0.5cm}
Extraordinary Sex Ratios: Cultural Effects on Ecological Consequences
}
\\
F. Moln\'{a}r Jr., T. Caraco, G. Korniss
\\
\end{flushleft}

\section*{Stationary solutions for a single female ratio}
\label{appendix_A}

When we consider a single female-ratio allele and a single male
mortality rate only, the model reduces to the mean-field equations
of Tainaka et al. [1]:
\begin{eqnarray}
\partial_{t} F & = & \theta \left(1-N\right) F M - \mu_{\rm f} F  \nonumber \\
\partial_{t} M & = & \left(1-\theta\right) \left(1-N\right) F M - \mu_{\rm m} M \;,
\label{eq:sa_fp}
\end{eqnarray}
where the total global density $N=F+M$.  For clarity we have dropped
subscripts for the resident group. These equations, in general, can
have three fixed points. One of these is the equilibrium at
extinction:
\begin{equation}
(F^{\rm o},M^{\rm o}) = (0,0) \label{sa_fp1}
\end{equation}
To obtain the non-trivial fixed points, we first manipulate the two
stationary state equations, Eqs.~(\ref{eq:sa_fp}), to write a simple
quadratic equation for the stationary total density,
\begin{equation}
N(1-N)=\frac{\mu_{\rm f}}{\theta} + \frac{\mu_{\rm m}}{1-\theta} \;,
\label{eq:sa_tot_fp}
\end{equation}
yielding solutions
\begin{equation}
N^{\pm} = \frac{1 \pm \sqrt{D}}{2} \label{N_tot_sa}
\end{equation}
with
\begin{equation}
D(\mu_{\rm f},\mu_{\rm m},\theta) = 1-4\left(\frac{\mu_{\rm
f}}{\theta} + \frac{\mu_{\rm m}}{1-\theta} \right) \;.
\label{D_sa_appendix}
\end{equation}
Finally, for the non-trivial female and male densities at
equilibrium, we have
%%%%%%%%%%%%%%%%%%%%%%%%%%%%%%%
\begin{equation}
(F^{\pm},M^{\pm}) = \left(\frac{\mu_{\rm m}}{1-\theta} \cdot
\frac{1}{1-N^{\pm}}\;  ,\;   \frac{\mu_{\rm f}}{\theta} \cdot
\frac{1}{1-N^{\pm}} \right) \;. \label{sa_fp23}
\end{equation}
%%%%%%%%%%%%%%%%%%%%%%%%%%%%%%%%

For $D\geq0$ all three fixed points are real.  The trivial (zero
density) solution [Eq.~(\ref{sa_fp1})] and the ``$+$" solution
[Eq.~(\ref{sa_fp23})] are locally stable, separated by an unstable
(saddle) fixed point, the ``$-$" solution in Eq.~(\ref{sa_fp23})
[the stability of these fixed points can be easily analyzed by
linearizing Eqs.~(\ref{eq:sa_fp})]. For $D<0$, however, only one
biologically meaningful (real) fixed point exists, the zero-density
solution [Eq.~(\ref{sa_fp1})], and extinction is always stable.

\begin{figure}[t]
\begin{centering}
\includegraphics[width=6in]{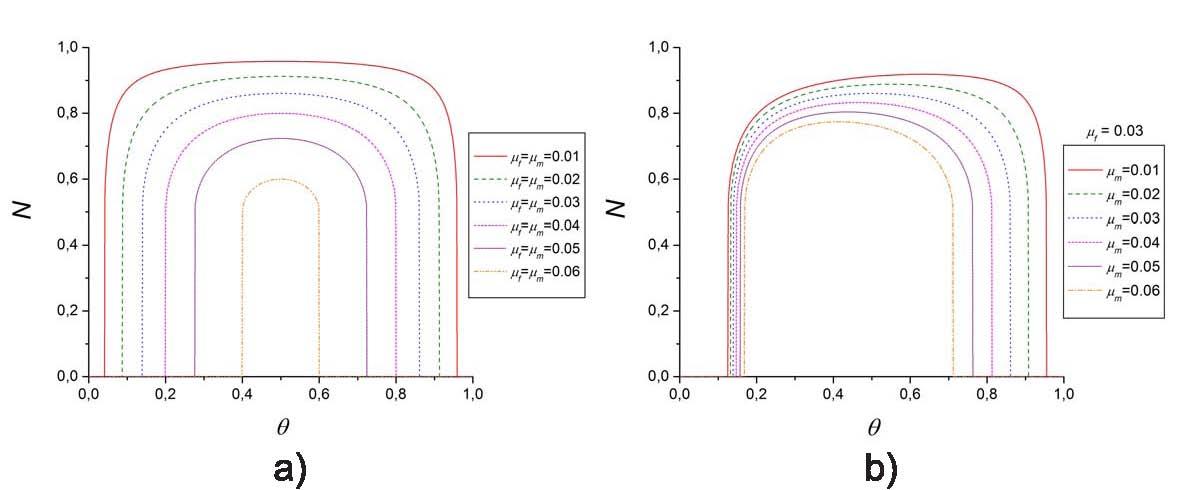}
\end{centering}
\caption{Stationary total population density in the single-allele
model [1] as a function of female ratio at birth,
for various mortality rates. (a) For identical female and male
mortality rates; (b) female and male mortality rates differ.}
\label{fig:N_vs_Theta}
\end{figure}

The biological significance of the structure of the above solutions
is two-fold [1]. First, for $D>0$, the system
exhibits the Allee effect.  Unless the (initial) population density
is sufficiently high ($N(0)>N^{-}$), the population goes extinct.
Second, provided that $\sqrt{\mu_{\rm f}} + \sqrt{\mu_{\rm m}}<1/2$,
there is a finite interval $\theta_{\rm c1}(\mu_{\rm f},\mu_{\rm
m})<\theta<\theta_{\rm c2}(\mu_{\rm f},\mu_{\rm m})$, where
$D(\mu_{\rm f},\mu_{\rm m},\theta)>0$, i.e., where the population
can persist at equilibrium (see Fig.~\ref{fig:N_vs_Theta}). These
boundaries, functions of the culturally transmitted mortality rate,
are given by:
%%%%%%%%%%%%%%%%%
\begin{equation}
\theta_{\rm c1,2}(\mu_{\rm f},\mu_{\rm m}) = \frac{(1+4\mu_{\rm
f}-4\mu_{\rm m})\pm \sqrt{(1+4\mu_{\rm f}-4\mu_{\rm m})^2 -
16\mu_{\rm f}}} {2} \;. \label{CultureConstraint1}
\end{equation}
%%%%%%%%%%%%%%%%
Between the two critical points, at
\begin{equation}
\theta^{*}=\frac{1}{1+\sqrt{\mu_{\rm m}/\mu_{\rm f}}} \;,
\end{equation}
total global density exhibits a maximum
\begin{equation}
N^{\max}=N^{+}(\theta^{*})=\frac{1+\sqrt{1-4(\sqrt{\mu_{\rm
f}}+\sqrt{\mu_{\rm m}})^2}}{2}
\end{equation}
where the female to male density ratio is
$F^{*}/M^{*}=\sqrt{\mu_{\rm m}/\mu_{\rm f}}$.

%%%%%%%%%%%%%%%%%%%%%%%%%%%%%%%%%%%%%%%%%%%%%%%%%%%%%%%%%%%%%%%%%%%%
%%%%%%%%%%%%%%%%%%%%%%%%%%%%%%%%%%%%%%%%%%%%%%%%%%%%%%%%%%%%%%%%%%%%%%

\section*{Stationary solutions for competing groups}
\label{appendix_B}

The four equations describing the competitive dynamics,
\begin{eqnarray}
\partial_{t} F_{1} & = & \theta_{1} \left(1-N\right) F_{1} \left(M_{1}+M_{2}\right) - \mu_{\rm f} F_{1} \nonumber \\
\partial_{t} M_{1} & = & \left(1-N\right) \left[\left(1-\theta_{1}\right) F_{1} \left(M_{1}+\frac{M_{2}}{2}\right)+\left(1-\theta_{2}\right) F_{2}  \left(\frac{M_{1}}{2}\right)\right] - \mu_{1}M_{1} \nonumber \\
\partial_{t} F_{2} & = & \theta_{2} \left(1-N\right) F_{2} \left(M_{1}+M_{2}\right) - \mu_{\rm f} F_{2} \nonumber \\
\partial_{t} M_{2} & = & \left(1-N\right) \left[ \left( 1-\theta_{2} \right) F_{2} \left(\frac{M_{1}}{2}+M_{2}\right) + \left(1-\theta_{1}\right) F_{1}  \left(\frac{M_{2}}{2}\right) \right] - \mu_{2} M_{2} \;,
\label{eq:2a_fp}
\end{eqnarray}
are considerably more complex than those of the single female-ratio
case.  Some of the fixed points, however, directly reflect those of
the single-allele case.  We obtained the remainder algebraically. Of
course, the competitive dynamics has the trivial fixed point where
all densities vanish.

\subsection*{Type-I fixed points: genetic and cultural exclusion}
\label{appendix_B_I}

The system admits stationary solutions where one allele and the
associated cultural trait are excluded, and the other allele has two
non-zero fixed points: the ``$-$" solutions in Eq.~(\ref{sa_fp23})
are always unstable, while the ``$+$" solutions are stable if the male
mortality rate of the allele with the higher female ratio is less
then twice the male mortality rate of the other allele. For example,
when allele 1 is excluded, these two fixed points are
%%%%%%%%%%%%%%%%%%%%%%%%%%%%%%%%%%%%%%%%%%%%%%%%%%%%%%%%
\begin{eqnarray}
(F_1^{\pm},M_1^{\pm}) & = & \left(0 \; , \; 0 \right)  \nonumber \\
(F_2^{\pm},M_2^{\pm}) & = & \left(\frac{\mu_2}{1-\theta_2} \cdot
\frac{1}{1-N^{\pm}} \; ,\; \frac{\mu_{\rm f}}{\theta_2} \cdot
\frac{1}{1-N^{\pm}} \right) \;,
\label{2a_fp_I}
\end{eqnarray}
%%%%%%%%%%%%%%%%%%%%%%%%%%%%%%%%%%%%%%%%%%%%%%%%%%%%%%%%
where $N^{\pm}$ is given by Eqs.~(\ref{N_tot_sa}) and
(\ref{D_sa_appendix}) with $\mu_{\rm m}=\mu_2$ and
$\theta=\theta_2$. Further, the ``$+$" solution above is stable
provided that $\mu_2/\mu_1<2$ and $\theta_2>\theta_1$. The stability
of these fixed points was checked numerically.

Considering the two symmetric cases by interchanging the extant
allele/culture with that excluded, we have four fixed points of this
type.  We refer to stable fixed points of this type [i.e., where the
densities of the extant allele are given by the ``$+$" solution in
Eqs.~(\ref{2a_fp_I})] as type-I fixed points.

\subsection*{Type-II fixed points: cultural coexistence}
\label{appendix_B_II}

Next, we obtained fixed points {\em not} related to those of the
single-allele dynamics. Numerical exploration never revealed
coexistence of both alleles and both cultural trait values.
However, we did find equilibrium populations with a single
female-ratio allele and a male cultural dimorphism.  Consider, e.g.,
$F_1=0$. The remaining  equations for the stationary state then
become
\begin{eqnarray}
0 & = & \left(1-N\right) \left(1-\theta_{2}\right) F_{2}  \left(\frac{M_{1}}{2}\right) - \mu_{1}M_{1}  \nonumber \\
0 & = & \theta_{2} \left(1-N\right) F_{2} \left(M_{1}+M_{2}\right) - \mu_{\rm f} F_{2}  \nonumber \\
0 & = & \left(1-N\right) \left( 1-\theta_{2} \right) F_{2}
\left(\frac{M_{1}}{2}+M_{2}\right) - \mu_{2} M_{2} \;,
\label{eq:2a_fp_reduced}
\end{eqnarray}
where the overall density is now $N=M_1+M_2+F_2$. After some tedious
algebra, we again find a simple quadratic equation for the overall
density:
\begin{equation}
N(1-N)=\frac{\mu_{\rm f}}{\theta_2} + \frac{2\mu_1}{1-\theta_2}
\label{eq:2a_tot_fp} \;,
\end{equation}
which has the solutions
\begin{equation}
\tilde{N}^{\pm} = \frac{1 \pm \sqrt{\tilde{D}}}{2}
\end{equation}
with
\begin{equation}
\tilde{D}(\mu_{\rm f},\mu_1,\theta_2) = %D(\mu_{\rm f},2\mu_1,\theta_2) =
1-4\left(\frac{\mu_{\rm f}}{\theta_2} + \frac{2\mu_1}{1-\theta_2}
\right) \;. \label{D_2a_appendix}
\end{equation}
The fixed points follow from Eqs.~(\ref{eq:2a_fp_reduced}) after
some further elementary manipulations
\begin{eqnarray}
(F_1^{\pm},M_1^{\pm}) & = & \left(0 \;,\; \frac{\mu_2/\mu_1-2}{\mu_2/\mu_1-1} \cdot \frac{\mu_{\rm f}}{\theta_2} \cdot \frac{1}{1-\tilde{N}^{\pm}}    \right)  \nonumber \\
(F_2^{\pm},M_2^{\pm}) & = & \left(\frac{2\mu_1}{1-\theta_2} \cdot
\frac{1}{1-\tilde{N}^{\pm}} \; ,\; \frac{1}{\mu_2/\mu_1-1} \cdot
\frac{\mu_{\rm f}}{\theta_2} \cdot \frac{1}{1-\tilde{N}^{\pm}}
\right) \label{2a_fp_II}
\end{eqnarray}
These fixed-point densities are biologically meaningful (real and
positive) if $\tilde{D}>0$ and $\mu_2/\mu_1>2$. The ``$-$" solution
above is always unstable, while the ``$+$" solution can be stable if
$\theta_2>\theta_1$ and a number of other necessary conditions are
satisfied (described below). We refer to stable fixed points given by
the ``$+$" solution in Eqs.~(\ref{2a_fp_II}) as type-II fixed
points of the two-allele system. The necessary conditions for
existence of these stable fixed points are $\sqrt{\mu_{\rm
f}}+\sqrt{2\mu_{1}}<1/2$ [from Eq.~(\ref{D_2a_appendix})] and
$\mu_2/\mu_1>2$ [from Eq.~(\ref{2a_fp_II})]. In this case, there is
a finite range of $\tilde{\theta}_{\rm c1}(\mu_{\rm
f},\mu_1)<\theta_2<\tilde{\theta}_{\rm c2}(\mu_{\rm f},\mu_1)$ where
$\tilde{D}(\mu_{\rm f},\mu_1,\theta_2)>0$, so that cultural
coexistence persists. The boundaries of this coexistence region are
given by:
%%%%%%%%%%%%%%%%%%
\begin{equation}
\tilde{\theta}_{\rm c1,2}(\mu_{\rm f},\mu_1) = %\theta_{\rm c1,2}(\mu_{\rm f},2\mu_1) =
\frac{(1+4\mu_{\rm f}-8\mu_1)\pm \sqrt{(1+4\mu_{\rm f}-8\mu_1)^2 -
16\mu_{\rm f}}} {2} \;.
\end{equation}
%%%%%%%%%%%%%%%%
Within this regime, the overall population density is maximal at
$\theta_2^*=1/(1+\sqrt{2\mu_1/\mu_{\rm f}})$ and the overall female
to male density ratio is $F_2^*/(M_1^*+M_2^*)=\sqrt{2\mu_1/\mu_{\rm
f}}$. The stability of these fixed points was checked numerically.
Interestingly, at the stable fixed point in Eq.~(\ref{2a_fp_II}) the
male density ratio is $M_1/M_2=\mu_2/\mu_1-2$; hence the relative
abundances of the male cultural trait values do not depend on the
female ratio.

Analogously, one can obtain fixed points of the same form as above
by choosing $F_2=0$ and simply interchanging indices $1$ and $2$ in
all respective expressions. Thus, combined, there are four fixed
points of this sort (consisting of both females and males of one
group and only males from the other group).

Considering all of the above, we have nine fixed points of
Eqs.~(\ref{eq:2a_fp}). Furthermore, a check with Mathematica [2]
assures that there are no other fixed points.

%{\color{red}
\section*{Asymmetric cultural transmission in cross-cultural mating}
\label{appendix_C}

One can generalize the homogeneous mean-field equations
(\ref{eq:2a_fp}) to capture asymmetry in the biparental transmission
of the cultural trait in males [3]. Cavalli-Sforza and
Feldman [4] point out that vertical
cultural-transmission probabilities can vary across different
combinations of parental phenotypes.  In our model, male offspring
resulting from mating of a female of group $i$ with a male of group
$j$ acquire the cultural trait of group $i$ or group $j$ with
probability $p$ and $q$, respectively ($p+q=1$). The corresponding
equations then read
%%%%%%%%%%%%%%%%%%%%%%%%%%%%%%%%%%%%%%%%%%%%%%%%%%%%%%%%%%%%%%%%%%%%
\begin{eqnarray}
\partial_{t} F_{1} & = & \theta_{1} \left(1-N\right) F_{1} \left(M_{1}+M_{2}\right) - \mu_{\rm f} F_{1} \nonumber \\
\partial_{t} M_{1} & = & \left(1-N\right) \left[\left(1-\theta_{1}\right) F_{1} \left(M_{1}+pM_{2}\right) + \left(1-\theta_{2}\right) qF_{2} M_{1}\right] - \mu_{1}M_{1} \nonumber \\
\partial_{t} F_{2} & = & \theta_{2} \left(1-N\right) F_{2} \left(M_{1}+M_{2}\right) - \mu_{\rm f} F_{2} \nonumber \\
\partial_{t} M_{2} & = & \left(1-N\right) \left[\left(1-\theta_{2}\right) F_{2} \left(pM_{1}+M_{2}\right) + \left(1-\theta_{1}\right) qF_{1} M_{2} \right] - \mu_{2} M_{2} \;.
\label{eq:2a_2s_asym}
\end{eqnarray}
The above equations allow for the same type-I fixed points as their
symmetric counterpart Eq.~(\ref{2a_fp_I}), with no change of the
form of the stable density of the surviving group (e.g., in complete
invasion/exclusion). The stability domain of this fixed point
changes however (see below).

Type-II fixed points, corresponding to cultural coexistence are also
possible, given by the solutions of the following equations (e.g.,
for $F_1$$=$$0$,  $M_1$$\neq$$0$, $F_2$$\neq$$0$, $M_2$$\neq$$0$),
%%%%%%%%%%%%%%%%%%%%%%%%%%%%%%%%%%%%%%%%%%%%%%%%%%%%%%%%%%%%%%%%%%%%%%%%%%
\begin{eqnarray}
0 & = & \left(1-N\right) \left(1-\theta_{2}\right) q F_{2}M_{1} - \mu_{1}M_{1}  \nonumber \\
0 & = & \theta_{2} \left(1-N\right) F_{2} \left(M_{1}+M_{2}\right) - \mu_{\rm f} F_{2}  \nonumber \\
0 & = & \left(1-N\right) \left( 1-\theta_{2} \right) F_{2} \left(pM_{1}+M_{2}\right) - \mu_{2} M_{2} \;.
\label{eq:2a_fp_reduced_asym}
\end{eqnarray}
The corresponding fixed points then become
\begin{eqnarray}
(F_1^{\pm},M_1^{\pm}) & = & \left(0 \;,\; \frac{\mu_2/\mu_1-1/q}{\mu_2/\mu_1-1} \cdot \frac{\mu_{\rm f}}{\theta_2} \cdot \frac{1}{1-\tilde{N}^{\pm}}    \right)  \nonumber \\
(F_2^{\pm},M_2^{\pm}) & = & \left(\frac{\mu_1/q}{1-\theta_2} \cdot
\frac{1}{1-\tilde{N}^{\pm}} \; ,\; \frac{p/q}{\mu_2/\mu_1-1} \cdot
\frac{\mu_{\rm f}}{\theta_2} \cdot \frac{1}{1-\tilde{N}^{\pm}} \right) \;,
\label{2a_fp_II_asym}
\end{eqnarray}
where $p=1-q$ and
\begin{equation}
\tilde{N}^{\pm} = \frac{1 \pm \sqrt{\tilde{D}}}{2}
\end{equation}
with
\begin{equation}
\tilde{D}(\mu_{\rm f},\mu_1,\theta_2,q) =
1-4\left(\frac{\mu_{\rm f}}{\theta_2} + \frac{\mu_1/q}{1-\theta_2} \right) \;.
\label{D_2a_appendix_asym}
\end{equation}
The above equations imply that type-II stable fixed point
(\ref{2a_fp_II_asym}) can only exist for $\sqrt{\mu_{\rm
f}}+\sqrt{\mu_{1}/q}<1/2$, $\mu_2/\mu_1>1/q$,  and
$\theta_2>\theta_1$, provided that $\tilde{\theta}_{\rm c1}(\mu_{\rm
f},\mu_1,q)<\theta_2<\tilde{\theta}_{\rm c2}(\mu_{\rm f},\mu_1,q)$,
%%%%%%%%%%%%%%%%%%
\begin{equation}
\tilde{\theta}_{\rm c1,2}(\mu_{\rm f},\mu_1,q) =
\frac{(1+4\mu_{\rm f}-4\mu_1/q)\pm \sqrt{(1+4\mu_{\rm f}-4\mu_1/q)^2 - 16\mu_{\rm f}}} {2} \;.
\end{equation}
%%%%%%%%%%%%%%%%
These conditions define the region where cultural coexistence
persists. In this region the ratio of coexisting cultures [male
density ratio for fixed point (\ref{2a_fp_II_asym})] is
$M_1/M_2=(q/p)(\mu_2/\mu_1-1/q)$.

\section*{References}

\begin{enumerate}
\item
Tainaka K, Hayashi T, Yoshimura J (2006) Sustainable sex ratio in
lattice populations. Europhys. Lett. 74: 554--559.
\item
Wolfram Research, Inc. (2010) Mathematica Version 8.0. Champaign,
IL: Wolfram Research, Inc.
\item
%\bibitem{referee}
This generalization was suggested by an anonymous referee during the
review process of this paper.
\item
%\bibitem{LCS_1981}
Cavalli-Sforza LL, Feldman MW (1981)
Cultural transmission and evolution: a quantitative approach.
Princeton: Princeton University Press.

\end{enumerate}

\end{document}